\documentclass[a4paper,fleqn,usenatbib]{mnras}
\usepackage[T1]{fontenc}
\usepackage{ae,aecompl}

\usepackage{graphicx}	
\usepackage{amssymb}	

\usepackage{multicol}
\usepackage[usenames]{color}
\usepackage{amsmath}
\usepackage{enumitem}
\usepackage{bm}

\newcommand {\bc}{\begin {center}}
\newcommand {\ec}{\end {center}}
\newcommand {\be}{\begin {equation}}
\newcommand {\ee}{\end {equation}}
\newcommand {\beq}{\begin {eqnarray}}
\newcommand {\eeq}{\end {eqnarray}}

\newcommand {\ergs}{{\rm erg\ \rm s^{-1}}}
\newcommand {\comment}[1]{}

\def\lbar {\lambda\hskip-5pt\raise3pt\hbox {--}}
\def\lbr {\lambda\raise2pt\hbox {\hskip-4pt{\scriptsize --}}_\C}

\renewcommand{\d}{{\rm d}}

\renewcommand{\d}{{\rm d}}

\title[Mass accretion rate pulsations in XRPs]
{
Spin-disc misalignment drives periodic accretion and pulse profile asymmetry in X-ray pulsars
}
\author[A. A.~Mushtukov et al.] 
{A.~A.~Mushtukov,$^{1,2}$\thanks{E-mail: a.mushtukov@ucl.ac.uk (AAM)}  
V.~Ganesh,$^{3}$
S.~S.~Tsygankov,$^{4}$
S.~Portegies Zwart,$^{5}$
A.~Palyam$^{6}$
\\ 
$^1$ Mullard Space Science Laboratory, University College London, Holmbury St. Mary, Surrey RH5 6NT, UK\\
$^2$ Astrophysics, Department of Physics, University of Oxford, Denys Wilkinson Building, Keble Road, Oxford OX1 3RH, UK\\
$^3$ California Institute of Technology, Pasadena, CA 91125, USA\\
$^4$ Department of Physics and Astronomy,  FI-20014 University of Turku, Finland \\
$^5$ Leiden Observatory, Leiden University, NL-2300RA Leiden, The Netherlands \\
$^6$ Mountain View High School, 3535 Truman Avenue, Mountain View, CA 94040, USA \\
} 

\pubyear{2026}

\begin{document}
\label{firstpage}
\pagerange{\pageref{firstpage}--\pageref{lastpage}}
\maketitle

\begin{abstract}
We investigate magnetospheric accretion in disc-fed X-ray pulsars assuming that the neutron star spin axis is not perpendicular to the disc plane. We focus on X-ray pulsars, where a geometrically thin disc is truncated far from the stellar surface
the channelled part of the accretion flow is expected to be guided by the large-scale dipolar magnetic field after coupling to it  near the disc--magnetosphere boundary.
Using numerical simulations of plasma motion from the inner disc edge to the neutron star surface, we show that a finite inclination between the disc normal and the stellar spin axis leads to periodic modulation of the mass accretion rate onto the magnetic poles even for a steady mass supply through the disc.
This purely geometrical effect arises because stellar rotation changes the orientation of the magnetosphere relative to the disc, producing phase-dependent mass loading of magnetic field lines. The amplitude and shape of the modulation are determined by the system geometry and by the ratio of the stellar spin period to the flow time through the magnetosphere.
The resulting variability affects the structure and luminosity of emitting regions near the neutron star surface and leads to asymmetric X-ray pulse profiles. Even without intrinsic asymmetries of the emission regions, this mechanism breaks the time-reversal symmetry expected for stationary accretion and naturally contributes to the observed asymmetry of pulse profiles and phase-resolved spectral features. The effect may also be relevant for ULX pulsars, where intrinsic accretion rate modulation can help preserve strong pulsations in the presence of geometric beaming.
\end{abstract}


\section{Introduction}
\label{sec:Intro}

\begin{figure*}
\centering 
\includegraphics[width=11.cm]{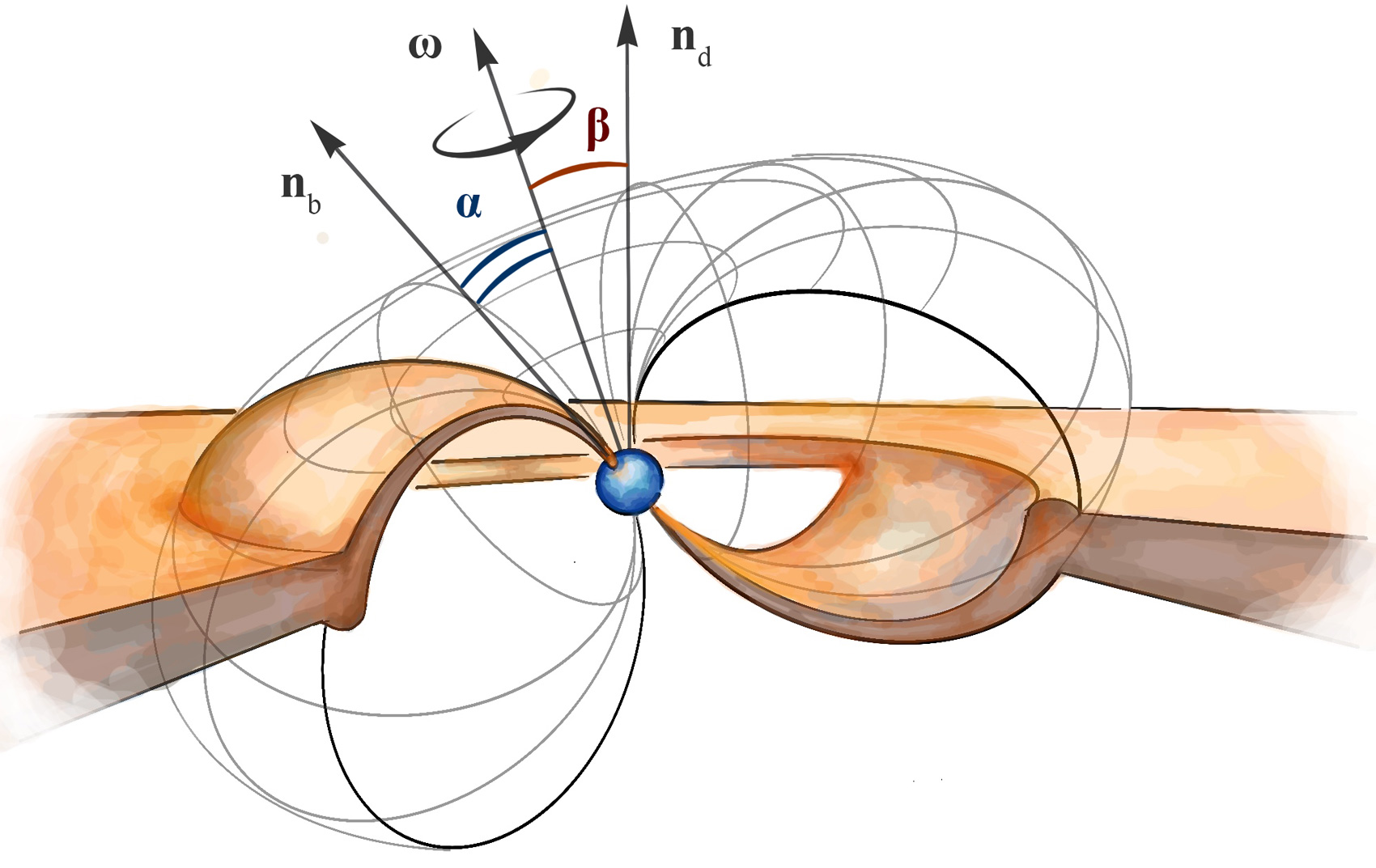}
\caption{
Schematic illustration of the accretion geometry in an XRP.
The magnetic field is assumed to be dominated by the dipole component, while the rotation axis is inclined with respect to the accretion disc plane.
The geometry of the NS rotation relative to the accretion disc plane is determined by the magnetic obliquity $\alpha$ (the angle between the rotational and magnetic field axes) and the angle $\beta$ between the normal to the accretion disc plane and the NS rotational axis.
}
\label{pic:scheme}
\end{figure*}

The presence of strongly magnetized neutron stars (NSs) in close binary systems results in the phenomenon of X-ray pulsars \citep[XRPs, see][for review]{2022arXiv220414185M}.
Magnetic field ($B$-field) strength at the surface of NSs in XRPs is typically of the order of $10^{12-13}\,{\rm G}$, which is confirmed by detection of cyclotron scattering features (see, e.g., \citealt{2019A&A...622A..61S}) that appear in X-ray spectra at energy $E_{\rm cyc}\simeq 11.6\,(B/10^{12}\,{\rm G})\,{\rm keV}$ due to resonant Compton scattering \citep{1986ApJ...309..362D}. 
A strong magnetic field affects elementary processes of radiation and matter interaction \citep{2006RPPh...69.2631H} and shapes the geometry of accretion flow in XRPs on spatial scales $\lesssim 10^8\,{\rm cm}$, where the accreting plasma couples to the magnetic field and is guided towards the magnetic poles (see Fig.~\ref{pic:scheme}).
The channelled flow then moves along magnetic field lines and reaches the NS surface in small regions near the magnetic poles.
The kinetic energy of the flow is converted into heat and is ultimately emitted in the form of X-ray photons and, at sufficiently high accretion rates, possibly neutrinos (see, e.g., \citealt{2018MNRAS.476.2867M,2025MNRAS.538.2396M}).
The misalignment between the magnetic and rotational axes of a NS leads to pulsations of the X-ray flux.
The pulse profile is determined by the geometry of NS rotation, the geometry of the emission region \citep{1976MNRAS.175..395B}, gravitational light bending \citep{1988ApJ...325..207R,2001ApJ...563..289K,2018MNRAS.474.5425M}, and the beam pattern of X-ray emission near the NS surface \citep{1973A&A....25..233G}. 
At mass accretion rates $\dot{M}\gtrsim 10^{18}\,{\rm g\,s^{-1}}$, the pulse profiles can also be affected by accretion flow covering the magnetosphere of a NS \citep{2017MNRAS.467.1202M,2023MNRAS.525.4176B} and by radiation-driven outflows from the accretion disc \citep{2009MNRAS.393L..41K,2021MNRAS.501.2424M,2023MNRAS.518.5457M}. 

The NS rotation parameters (NS inclination and magnetic obliquity), which are encoded in variations of the polarization angle over the spin period, have recently been probed in some XRPs by the Imaging X-ray Polarimetry Explorer (IXPE; e.g. \citealt{2021AJ....162..208S}), using pulse-phase–resolved polarization measurements and the rotating vector model \citep{2022NatAs...6.1433D,2022ApJ...941L..14T,2023A&A...675A..48T,2023MNRAS.524.2004M,2023A&A...675A..29M,2024A&A...691A.123P,2024NatAs...8.1047H}.

At the same time, interpretation of XRP pulse profiles remains challenging and usually requires additional assumptions about the beam pattern and magnetic-field geometry. 
Observed profiles are often markedly asymmetric. 
Such asymmetry is commonly attributed to non-antipodal emitting regions \citep{1996ApJ...467..794K,2010A&A...517A...8S,2011A&A...526A.131C,2012A&A...540A..35S}. 
Additional complexity can be introduced by the energy dependence of the beam pattern \citep{2011A&A...532A..76F}. 
In most of these interpretations, the mass accretion rate onto the NS surface is treated as an external input parameter.
This leaves open the question of whether the geometry of the disc--magnetosphere system itself can produce a periodic modulation of the accretion rate.

The interaction between a tilted stellar magnetosphere and an accretion disc has been studied for decades in the context of magnetically driven warping and precession of the inner disc \citep{1980A&A....86..192A,1980SvAL....6...14L,1999ApJ...524.1030L}. 
These works demonstrated that magnetic torques can warp the inner disc and drive its precession, while viscous stresses tend to reduce the tilt and promote alignment. 
Global 3D MHD simulations of magnetospheric accretion have also been performed for inclined dipoles \citep[e.g.][]{2003ApJ...595.1009R,2004ApJ...610..920R,2013MNRAS.430..699R,2016MNRAS.459.2354B}, although in most cases the stellar spin axis was assumed to be aligned with the disc normal.

\begin{figure*}
\centering 
\includegraphics[width=17.5cm]{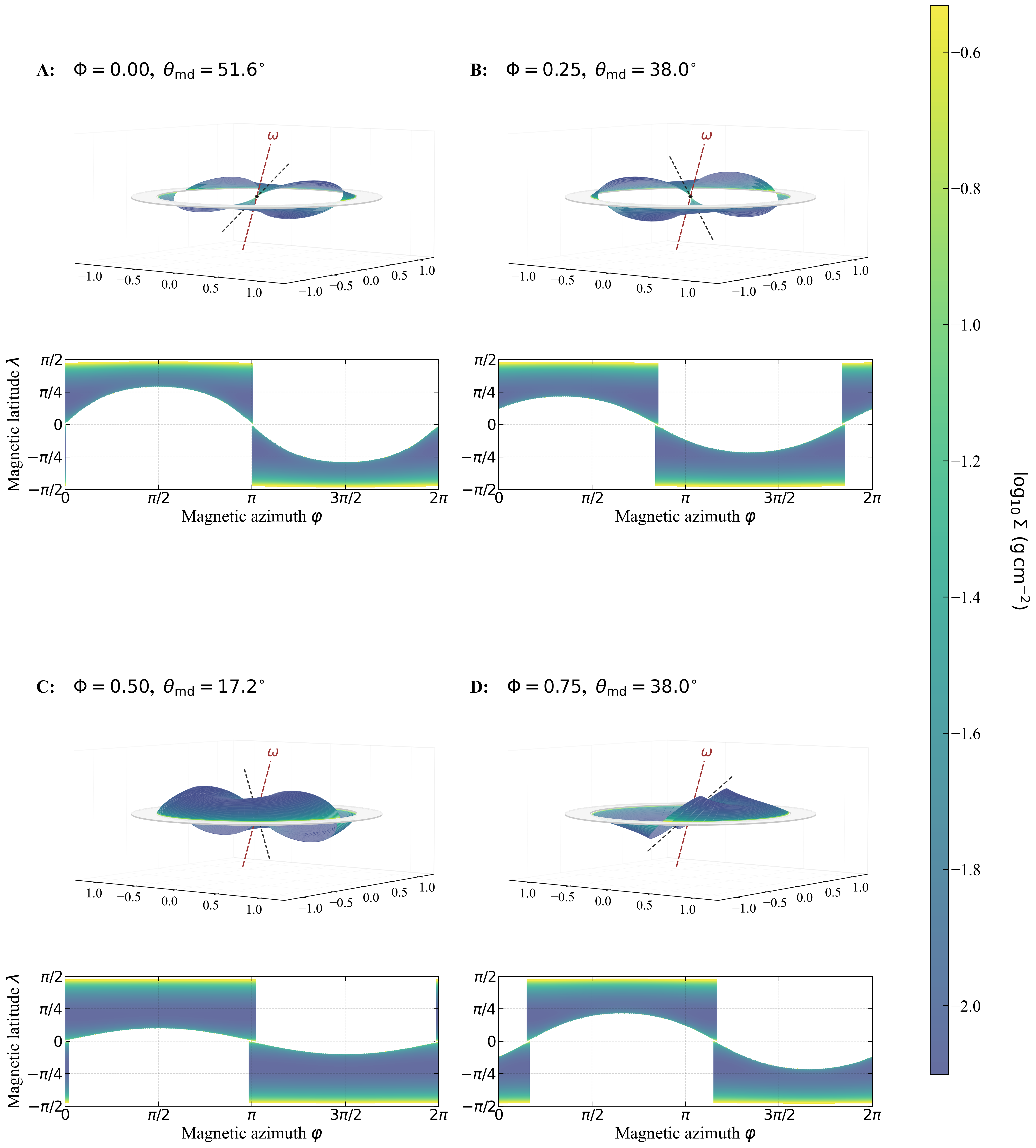}
\caption{
{Geometry of magnetically channelled accretion onto a NS at four rotational phases. The upper panels show the three-dimensional structure of the accretion flow, with the accretion disc (grey), NS (black), magnetic axis (dotted line), spin axis (dashed line), and the two magnetospheric accretion curtains. The colour of the flow indicates the local surface density of the accreting plasma. The lower panels show the corresponding distributions of the surface density, $\Sigma$, over the magnetospheric surface as a function of magnetic azimuth, $\varphi$, and magnetic latitude, $\lambda$. The white dashed curve marks the intersection of the disc mid-plane with the dipolar magnetic field, while the dotted curves indicate the finite thickness of the accretion disc. The adopted geometry corresponds to a magnetic obliquity of $\alpha=0.6$ rad and a misalignment angle between the spin axis and the disc normal of $\beta=0.3$ rad. The four panels correspond to rotational phases $\Phi=0.00$, $0.25$, $0.50$, and $0.75$.}
}
\label{pic:coverage}
\end{figure*}

{
The more general case, in which the stellar spin axis itself is inclined with respect to the disc, was investigated by \citet{2021MNRAS.506..372R} using global 3D MHD simulations. 
Their calculations showed that the tilted rotating magnetosphere can warp, tilt, thicken, and drive precession of the inner disc. 
In models where both the stellar spin axis and the magnetic axis were tilted, \citet{2021MNRAS.506..372R} also found variations of the accretion rate associated with the changing angle between the magnetic moment and the disc: funnel accretion was more efficient when the magnetosphere was more strongly inclined relative to the disc. 
However, in those simulations the accretion rate variability was embedded in the full nonlinear evolution of the disc--magnetosphere system and was therefore quasi-periodic rather than a clean phase-locked signal.
This leaves open a distinct question: can the same geometrical effect, isolated from the back reaction of the magnetosphere on the disc, produce a strictly periodic modulation of the accretion rate in the large-magnetosphere regime relevant to classical X-ray pulsars? 
This question is important because a phase-locked modulation of the mass supply to the magnetic poles can affect the observed X-ray pulse profile, whereas a purely quasi-periodic or stochastic modulation would mainly contribute additional noise.
}

In classical XRPs, by contrast, the disc is expected to be geometrically thin and truncated at radii much larger than $R_{\rm NS}$.
In this regime, after coupling to the magnetic field near the disc--magnetosphere boundary, the plasma motion is expected to be largely controlled by the large-scale dipolar field, and geometrical effects can be considered separately from the full MHD disc--magnetosphere interaction.
It is therefore unclear whether the variability found in global simulations with compact magnetospheres can be directly extrapolated to the large-magnetosphere regime of XRPs.

A finite spin--disc misalignment is not expected to be long-lived in persistently accreting systems, where accretion torques tend to align the NS spin axis with the angular-momentum axis of the inflowing matter on the spin-up time-scale \citep{2021MNRAS.505.1775B}. 
This argument is less restrictive for Be-XRPs (see, e.g., \citealt{2011Ap&SS.332....1R}), where accretion is often episodic. 
In such systems, the effective alignment time-scale can be significantly longer, and may be further extended if the angular-momentum axis of the accreted material varies between outbursts. 
Numerical studies of Be/X-ray binaries indeed show that the Be-star decretion disc can be substantially misaligned with the binary plane and that such a geometry can strongly affect the mass-transfer pattern during outbursts \citep{2013PASJ...65...41O,2014ApJ...790L..34M}. 
These considerations provide a motivation to examine configurations with a small but finite spin--disc misalignment.

An additional and independent motivation is provided by the observed super-orbital variability in a number of XRPs.
Such variability, on time-scales much longer than the orbital period, is commonly interpreted either in terms of a warped and precessing accretion disc or precession of a NS itself (see, e.g., \citealt{1976ApJ...209..562G,2000ApJ...539..392S,2009A&A...494.1025S,2013A&A...550A.110S}). 
In both scenarios, the mutual orientation of the disc and the NS rotation axis varies with time, implying the presence of a non-zero angle between the spin axis and the disc normal. 
This may lead to a periodically changing geometry of the disc--magnetosphere interaction and may therefore modulate the mass accretion rate onto the NS surface.

An illustrative example is Her X-1, where the $\sim 35$-d super-orbital cycle is attributed to a tilted and precessing accretion disc, while the pulse profile evolves systematically over the same cycle \citep{1976ApJ...209..562G,2000ApJ...539..392S,2009A&A...494.1025S,2013A&A...550A.110S}. 
Although these variations are often interpreted in terms of variable screening by the inner disc, the changes in the relative orientation of the disc and the NS spin may produce reproducible modifications of the observed pulse morphology.

In this paper, we consider classical XRPs, i.e., we focus on the regime of a geometrically thin disc and a large magnetosphere. 
After the plasma couples to the magnetic field near the disc--magnetosphere boundary, we prescribe its motion along dipolar field lines in a rigid stellar magnetosphere.
We study accretion flow motion from the inner edge of a disc to the NS surface for a non-zero angle between the disc normal and the stellar spin axis, while neglecting the back reaction of the magnetosphere on the disc.
This simplified setup is complementary to global 3D MHD simulations: it does not follow the warping, thickening, or precession of the inner disc, but instead isolates the geometrical phase dependence of mass loading in the large-magnetosphere limit.
In particular, we ask whether the geometrical tendency seen in the 3D MHD calculations of \citet{2021MNRAS.506..372R} can become a clean, phase-locked signal when the long-term dynamical response of the disc is removed from the problem.
Our aim is to determine whether this purely geometrical misalignment can produce a coherent spin-periodic modulation of the mass accretion rate under stationary mass supply through the disc, and to explore its implications for the formation and asymmetry of X-ray pulse profiles.

\section{Model set up}
\label{sec:Model}

\subsection{Geometry of the accretion flow}

In this paper, we examine the case of disc-fed accretion, where plasma is channeled from the inner edge of the disc on to the magnetic poles of a NS. 
We consider the case in which the stellar spin axis is not exactly perpendicular to the disc plane. 
The angle between the spin axis and the disc normal is denoted by $\beta$, while the angle between the spin axis and the magnetic dipole axis is denoted by $\alpha$ (see Fig.~\ref{pic:scheme}).

The geometrical consequences of this configuration are illustrated in Fig.~\ref{pic:coverage}, which shows the magnetically channelled accretion flow at four rotational phases. 
As the NS rotates, the orientation of the dipolar magnetosphere with respect to the accretion disc changes periodically, shifting the location where the disc intersects the magnetic field. 
Consequently, the mass loading of individual magnetic field lines becomes phase dependent, leading to a time-dependent distribution of plasma over the magnetosphere even for a constant mass supply through the disc.

The disc is truncated at the magnetospheric radius determined by the mass accretion rate and the NS magnetic field strength and structure. 
If the $B$-field is dominated by the dipolar component, the magnetospheric radius can be estimated as
\beq\label{eq:Rm}
R_{\rm m} \approx 1.8\times 10^8\,\Lambda B_{12}^{4/7}\dot{M}_{17}^{-2/7}m^{-1/7}R_6^{12/7}\,\,{\rm cm},
\eeq
where $\Lambda<1$ is a factor that accounts for the geometry of the inflow (see, e.g., Chapter 6.3 in \citealt{2002apa..book.....F}, and \citealt{2019A&A...626A..18C} for discussion at high $\dot M$), $B_{12}$ is the surface magnetic field in $10^{12}\,{\rm G}$, $\dot M_{17}$ is the mass accretion rate in units of $10^{17}\,{\rm g\,s^{-1}}$, $m$ is the NS mass in $M_\odot$, and $R_6$ is the stellar radius in $10^6\,{\rm cm}$.

The effective temperature of the accretion disc at a radius $r$ can be estimated as
\beq\label{eq:Teff_1}
\sigma _{\rm SB}T^4_{\rm eff}=
\frac{3}{8\pi}\frac{GM\dot{M}}{r^3}
\left[1-\beta_{\rm vis}\left(\frac{R_{\rm m}}{r}\right)^{1/2}\right],
\eeq
where $\sigma_{\rm SB}$ is the Stefan-Boltzmann constant, and $\beta_{\rm vis}\in[0;1]$ parametrizes the position where viscous stress vanishes \citep{2017A&A...608A..17T}. 
From equations~(\ref{eq:Rm}) and (\ref{eq:Teff_1}), the disc temperature near the truncation radius is estimated as
\beq\label{eq:Teff_2}
T_{\rm eff}\lesssim 
0.03\,B_{12}^{-3/7}\dot{M}_{17}^{13/28}\,\,\,
{\rm keV}.
\eeq
The actual plasma temperature in the coupling region between the disc and magnetosphere may be higher, since equations~(\ref{eq:Teff_1}) and (\ref{eq:Teff_2}) do not include the effects of disc--magnetosphere interaction.

{
At the relatively low accretion rates considered here, the disc is expected to be truncated in the gas-pressure-dominated region, where the opacity is mainly determined by free--free absorption. 
Following the same prescription as in \citet{2026arXiv260710342M}, the geometrical half-thickness of the disc at the magnetospheric radius can be estimated as
\beq
H_{\rm d} \simeq
1.2\times10^{6}\,
\alpha_{\rm d}^{-1/10}
L_{37}^{3/20}
m^{-21/40}
R_{6}^{3/20}
R_{{\rm m},8}^{9/8}
\ {\rm cm},
\eeq
where $\alpha_{\rm d}$ is the Shakura-Sunyaev viscosity parameter and $R_{{\rm m},8}=R_{\rm m}/10^{8}\,{\rm cm}$ (see, \citealt{2007ARep...51..549S} for details). 
Thus,
\beq
\frac{H_{\rm d}}{R_{\rm m}}
\simeq
10^{-2}\,
\alpha_{\rm d}^{-1/10}
L_{37}^{3/20}
m^{-21/40}
R_{6}^{3/20}
R_{{\rm m},8}^{1/8}
\ll 1,
\eeq
which justifies treating the disc as geometrically thin in the present calculations.
}

We do not attempt to describe the full plasma dynamics inside a closed magnetosphere or the detailed penetration of matter through the disc--magnetosphere boundary. Instead, we assume that, after coupling to the stellar magnetic field in the disc--magnetosphere interaction region, the channelled part of the accretion flow moves along prescribed magnetic field lines.
We consider a pure dipole magnetic field and neglect field distortion caused by the interaction between the NS magnetosphere and accretion disc.
In spherical coordinates $(r,\theta,\varphi)$, a dipole field line can be represented as
\beq\label{eq:dip_fl}
r=R_{\rm max}\cos^2 \lambda=R_{\rm max}\sin^2 \theta, 
\eeq
where $R_{\rm max}$ sets the scale of the field line, $\lambda$ is latitude, and $\theta=\pi/2-\lambda$ is the co-latitude. 

For a dipole aligned with the disc normal, all field lines intersecting the disc at its inner edge have the same scale, $R_{\rm max}=R_{\rm m}$. 
For an inclined dipole, the field line scale depends on azimuth because the line of intersection between the dipolar surface and the disc plane is no longer axisymmetric. 
In the case $\beta=0$, this dependence is stationary in the rotating frame. 
For $\beta\neq 0$, the orientation of the dipole with respect to the disc varies with rotational phase, and the location where a given magnetic meridian intersects the disc becomes explicitly time-dependent.
As a result, even for a constant mass supply through the disc, the mass loading of magnetic field lines becomes phase-dependent.
This time-dependent mass loading is the key ingredient of the present model.

In the numerical calculations, the magnetosphere is represented by a set of dipolar field lines labelled by the azimuthal coordinate $\phi$, and the motion of the channelled plasma component is followed along each field line in terms of the coordinate $\lambda$.
Matter is injected in the vicinity of the local disc--field intersection and then moves towards one of the magnetic poles. 
This prescription should be understood as an effective description of the magnetically guided part of the flow, rather than as a self-consistent calculation of plasma loading and penetration through the magnetospheric boundary.
The injection region is therefore time-dependent for $\beta\neq0$, which directly leads to a modulation of the accretion flow reaching the NS surface.

The surface element on the dipole is expressed as
\beq
\d S=R_{\rm max}^2 \cos^4\lambda (1+3\sin^2\lambda)^{1/2} \d\lambda\d\varphi,
\eeq
while the arc length along a field line satisfies
\beq
\d x = R_{\rm max}\cos\lambda (1+3\sin^2\lambda)^{1/2} \d\lambda.
\eeq
The angle between the local radial vector and the tangent to the dipole line is
\beq\label{eq:chi}
\chi={\rm atan}[0.5\,{\rm tan}^{-1}\lambda].
\eeq

Following \citet{2024MNRAS.530..730M}, we construct magnetospheric maps of the local surface density, accretion flow velocity, and acceleration along the field lines. 
Since the numerical scheme itself has been described in detail in \citet{2024MNRAS.530..730M}, here we restrict ourselves to the aspects specific to the misaligned-rotator geometry.

\subsection{Dynamics of the accretion flow}
\label{sec:Dynamics}

We assume that the channelled component of the accretion flow starts its motion along B-field lines after coupling to the magnetosphere near the inner disc radius with the initial velocity $v_i$.
Within this effective field-guided description, plasma motion along the magnetic field lines is governed by the projection of gravity and centrifugal force on to the field direction. In the general formulation, radiative force can also be included.
Gas pressure gradients are assumed to be small and neglected. 
We focus on relatively low-luminosity XRPs, $L_{\rm X}\lesssim 10^{37}\,\ergs$, and therefore neglect the radiative force in the baseline model.

The gravitational acceleration projected on to a field line is
\beq
\left|a_{{\rm grav},||}\right|=\cos\chi\,\frac{GM}{r^2}\simeq 1.328\times 10^{10}\,m r_8^{-2}\cos\chi \,\,{\rm cm\,s^{-2}}.
\eeq
It changes sign across the equator: positive at $\lambda>0$ and negative at $\lambda<0$. Its magnitude increases towards the NS.

In the frame co-rotating with the NS, the centrifugal force contributes to the field-aligned acceleration,
\beq
a_{\rm cen,\parallel}
= - \mathbf n_B \cdot
\left[
\boldsymbol\Omega
\times (
\boldsymbol\Omega \times \mathbf r ) \right],
\eeq
where $\mathbf n_B$ is the unit tangent vector to the magnetic field line.
For $\beta\neq0$, this contribution depends explicitly on rotational phase because the magnetic field co-rotates with the star while the disc
defines a fixed external reference plane.
As a result, the total acceleration along a given field line is time-dependent even for a stationary mass supply.

Under the assumptions of the model, the channelled component of the flow moves strictly along field lines, and the one-dimensional dynamics is described by the projection of external forces on to the field direction. 

A further important ingredient of the model is the treatment of the disc--magnetosphere interface. 
Whenever matter returns to the disc-interaction region, its subsequent motion inside the disc is not followed explicitly. 
Instead, it is redistributed over the magnetic meridians intersecting the disc and re-injected into the magnetosphere with a prescribed initial velocity. 
This effective prescription, inherited from \citet{2024MNRAS.530..730M}, is intended to mimic matter redistribution near the inner disc edge while keeping the problem computationally tractable. 
In the baseline model, the external mass supply through the disc is taken to be constant.
In construction of a numerical model, we follow the ideas described in \citet{2024MNRAS.530..730M} (see also Appendix~\ref{app:NumMod}).

\section{Results of numerical simulations}
\label{sec:NumRes}

\subsection{Pulsating mass accretion rate at the NS surface}

\subsubsection{Constant mass accretion rate at $R_{\rm m}$}

\begin{figure}
\centering 
\includegraphics[width=8.7cm]{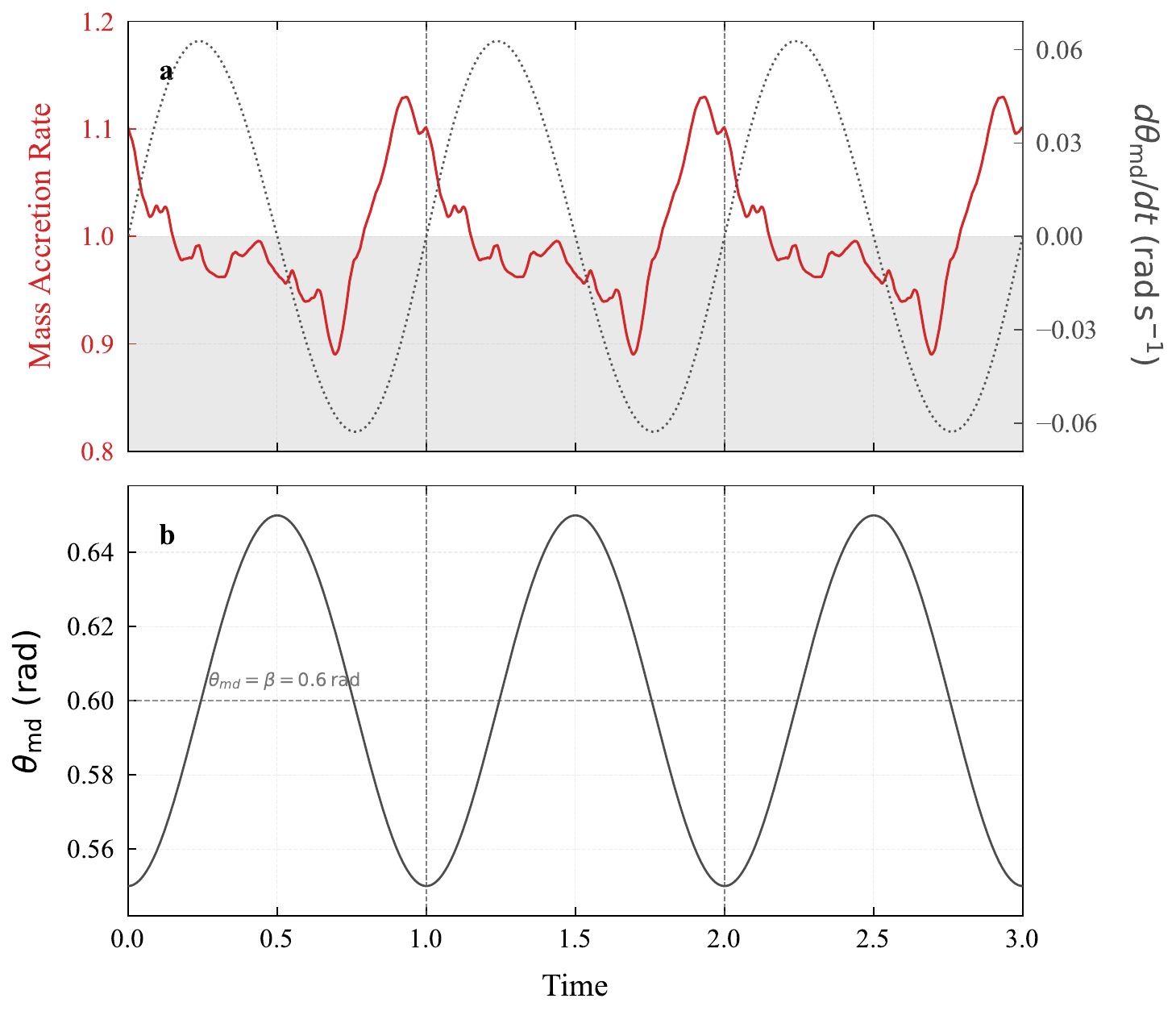}
\caption{
Geometrical origin of the periodic modulation of the mass accretion rate for the fiducial model with $\alpha=0.6$\,rad, $\beta=0.05$\,rad, $P=5$\,s, $R_{\rm m}=2\times10^{8}$\,cm, and $v_{\rm i}=10^{7}\,{\rm cm\,s^{-1}}$.
The upper panel shows the mass accretion rate at the NS surface, normalized to the constant mass supply at the inner disc radius (red solid curve), together with the time derivative of the instantaneous angle $\theta_{\rm md}$ between the magnetic axis and the disc normal (grey dotted curve; right vertical axis).
The lower panel shows the instantaneous angle $\theta_{\rm md}$.
The horizontal dotted line marks the magnetic obliquity, $\theta_{\rm md}=\alpha=0.6$\,rad.
Time is given in units of the NS spin period.
The accretion rate modulation follows the changing disc--magnetosphere orientation with a finite phase delay, comparable to the characteristic propagation time of the flow from the inner disc edge to the NS surface.
}
\label{pic:flow_00}
\end{figure}

\begin{figure*}
\centering 
\includegraphics[width=16.cm]{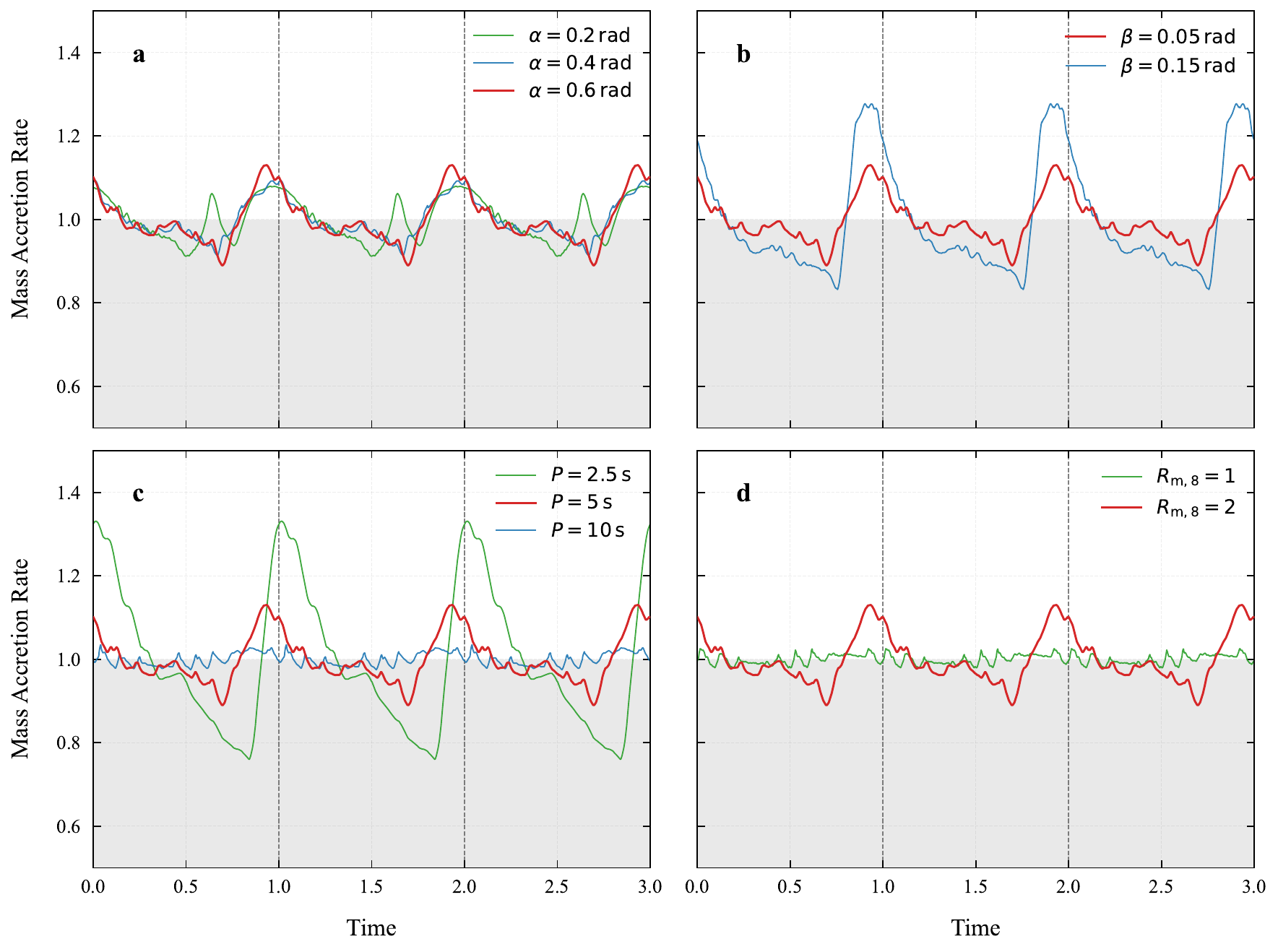}
\caption{
Mass accretion rate at the NS surface (in units of mass accretion rate at the inner disc radius) as a function of time given in units of NS spin period.
Different panels illustrate influence of different parameters of a problem:
(a) angle between NS rotation axis and magnetic axis,
(b) angle between NS rotation axis and normal to accretion disc plane, 
(c) spin period of a NS, 
(d) inner radius of accretion disc.
Red line in all panels shows result for the fiducial parameters: 
$\alpha = 0.6\,{\rm rad}$, 
$\beta = 0.05\,{\rm rad}$,
$P=5\,{\rm s}$, 
$R_{\rm m}=2\times 10^8\,{\rm cm}$, 
$v_{\rm i}=10^7\,{\rm cm\,s^{-1}}$.
The small-scale step-like variations are caused by the finite number of azimuthal cells used in the numerical calculation.
}
\label{pic:flow_0}
\end{figure*}

The results of the numerical simulations are summarised in Fig.~\ref{pic:flow_0}, which shows the mass accretion rate at the NS surface as a function of time in units of NS spin period. 
The accretion rate is normalised to the mass accretion rate at the inner disc radius, which is assumed to be constant here. 
Therefore, all variability of the accretion rate at the stellar surface arises solely from the geometrical effect associated with the spin--disc misalignment.
The small-scale step-like structure visible in the curves is a numerical effect caused by the finite azimuthal discretisation of the magnetospheric flow. The calculations shown in Fig.~\ref{pic:flow_0} were performed using $N_{\phi}=120$ azimuthal cells and approximately $N_{\lambda}=400$ cells along the magnetic field lines. 
The large-scale periodic modulation discussed below is not associated with this discretisation effect.

To investigate the influence of different model parameters on the variability of the mass accretion rate at the NS surface, we first simulate a fiducial case (see red line in Fig.~\ref{pic:flow_0}) using the following parameters: 
$\alpha = 0.6\,{\rm rad}$,
$\beta = 0.05\,{\rm rad}$,
$P = 5\,{\rm s}$,
$R_{\rm m}=2\times 10^8\,{\rm cm}$,
and 
$v_i=10^7\,{\rm cm\,s^{-1}}$.
{
The geometrical origin of the modulation in the fiducial model is illustrated in Fig.~\ref{pic:flow_00}.
During NS rotation, the instantaneous angle $\theta_{\rm md}$ between the magnetic axis and the disc normal varies periodically around the magnetic obliquity $\alpha$.
The grey dotted curve in the upper panel shows the time derivative $d\theta_{\rm md}/dt$, which characterizes how rapidly the orientation of the magnetosphere relative to the disc changes.
The mass accretion rate at the NS surface is clearly correlated with these geometrical variations, but responds with a finite phase delay.
This delay naturally arises because matter loaded onto magnetic field lines near the inner disc radius requires a finite time to propagate through the magnetosphere before reaching the stellar surface.
For the fiducial parameters, this delay is comparable to the characteristic propagation time,
\beq
t_{\rm flow}\simeq
\pi
\left(
\frac{R_{\rm m}^{3}}
{2GM}
\right)^{1/2}\sim 0.5\,{\rm s},
\eeq
demonstrating that the observed modulation reflects both the instantaneous geometry of the disc--magnetosphere system and the finite travel time of the accreting plasma.
}

Fig.~\ref{pic:flow_0}a illustrates the dependence of the accretion rate modulation on the angle $\alpha$ between the stellar spin axis and the magnetic dipole axis. 
Variations of $\alpha$ have only a weak effect on the amplitude of the oscillations.
However, the rotational phase of the accretion rate maximum depends noticeably on $\alpha$.
Thus, the magnetic obliquity $\alpha$ primarily shifts the phase of the modulation without significantly affecting its amplitude.

Fig.~\ref{pic:flow_0}b demonstrates the influence of the angle $\beta$ between the disc normal and NS spin axis. 
This parameter strongly affects the amplitude of mass accretion rate variability. 
As the spin--disc misalignment increases, the amplitude of the accretion rate oscillations becomes significantly larger. 
In our simulations the modulation amplitude reaches $\sim20$--$40$ per cent of the mean accretion rate, despite the fact that the mass supply through the disc remains constant.

The dependence on the NS spin period is shown in Fig.~\ref{pic:flow_0}c: the modulation amplitude decreases with increasing spin period. 
Finally, Fig.~\ref{pic:flow_0}d shows the effect of the magnetospheric radius. 
The simulations correspond to magnetospheric radii of order $R_{\rm m}\sim10^{8}\,\mathrm{cm}$. 
The variability amplitude increases with increasing $R_{\rm m}$.
These trends can be summarised as follows: the modulation amplitude increases with increasing misalignment angle $\beta$ and magnetospheric radius $R_{\rm m}$, and decreases with increasing spin period $P$.
At a qualitative level, these trends are consistent with the idea that the variability amplitude depends on how strongly the magnetosphere changes its orientation with respect to the disc during the characteristic propagation time of the flow from the inner disc edge to the stellar surface, $t\simeq \pi (R_{\rm m}^3/2GM)^{1/2}\approx 0.2\,R_{\rm m,8}^{3/2}m^{-1/2}\,{\rm s}$. 
For shorter spin periods or larger magnetospheric radii, the rotating misaligned magnetosphere can change its orientation more noticeably during this time, which may lead to a more efficient phase-dependent modulation of the mass loading of magnetic field lines and, consequently, to a larger variability amplitude.

\subsubsection{Fluctuating mass accretion rate at $R_{\rm m}$}

\begin{figure}
\centering 
\includegraphics[width=8.5cm]{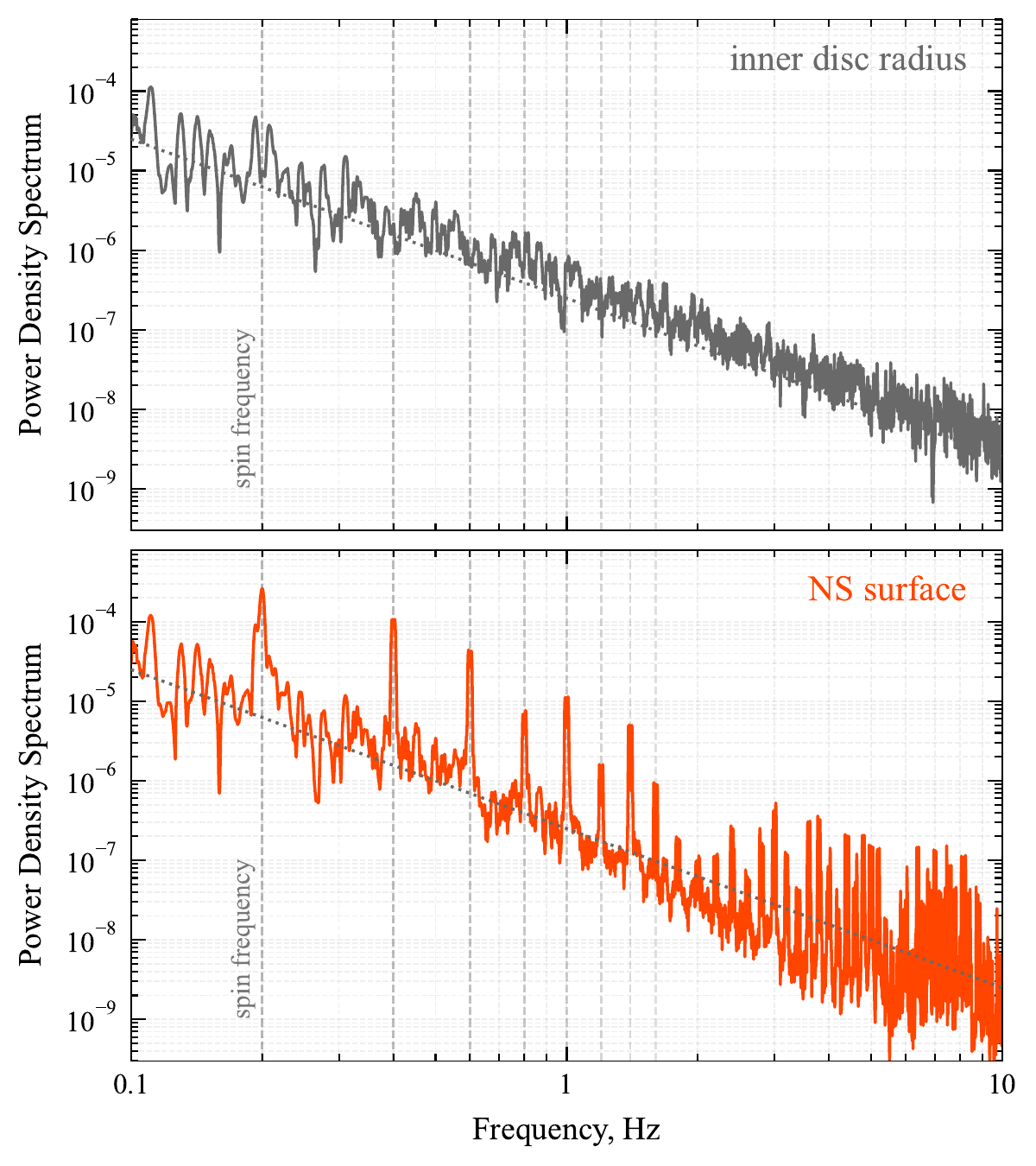}
\caption{
Generic PDS of the mass accretion rate at the inner disc radius (upper panel) and PDS of the mass accretion rate at the NS surface (lower panel).
The latter shows peaks at the spin frequency of a NS and a number of harmonics. 
Parameters:
$\alpha=0.6\,{\rm rad}$, 
$\beta=0.05\,{\rm rad}$, 
$R_{\rm m}=2\times 10^8\,{\rm cm}$, 
$P=5\,{\rm s}$, 
$v_{i}=10^7\,{\rm cm\,s^{-1}}$.
}
\label{pic:flow_PDS}
\end{figure}

Mass accretion rate at the inner radius of accretion disc in XRPs is expected to fluctuate due to the stochastic nature of viscosity in the disc \citep{1997MNRAS.292..679L,2009A&A...507.1211R,2019MNRAS.486.4061M}.
To investigate how such variability propagates through the magnetosphere, we impose a time-dependent mass accretion rate at the inner disc radius.
The input time series is generated using the algorithm of \citet{1995A&A...300..707T}, assuming a prescribed power density spectrum (PDS) and fractional rms.
In the simulations presented here, we adopt a power-law PDS of the form $P(f) \propto f^{-2}$.
The resulting stochastic accretion rate is used as an outer boundary condition for the magnetospheric flow (see Appendix~\ref{sec:AppPDS}).
Power spectra shown in Fig.~\ref{pic:flow_PDS} were computed from a 230-s realisation for the fiducial model with NS spin period $P=5,{\rm s}$, $\alpha=0.6,{\rm rad}$, $\beta=0.05,{\rm rad}$,
$R_{\rm m}=2\times 10^8,{\rm cm}$, and initial velocity of accretion flow at the inner disc radius
$v_{i}=10^7\,{\rm cm,s^{-1}}$.

The effect of stochastic variability of the mass accretion rate supplied by the disc is illustrated in Fig.~\ref{pic:flow_PDS}.
The upper panel shows the PDS of the input accretion rate fluctuations, characterised by a red-noise component with a power-law shape.
After propagation through the magnetosphere, the broadband variability is largely preserved: the slope of the power spectrum at low frequencies remains close to that imposed at the disc--magnetosphere boundary.
We stress, however, that this conclusion is based on a single representative realisation and does not constitute a systematic exploration of parameter space.
Within the considered setup, we do not find strong evidence for a significant distortion of the large-scale stochastic variability by the magnetospheric flow.
In particular, no clear suppression of variability is seen at frequencies higher than the NS spin frequency.

At the same time, the PDS of the accretion rate at the NS surface (lower panel in Fig.~\ref{pic:flow_PDS}) exhibits additional narrow features at the NS spin frequency and its harmonics.
These peaks arise due to periodic modulation of the accretion flow by the rotating magnetosphere.
Thus, the magnetosphere preserves the broadband stochastic variability while superimposing a coherent signal at the spin frequency and its harmonics.
In this sense, the magnetosphere acts as a transfer medium that transmits the input fluctuations while imprinting a periodic modulation associated with the NS rotation.

\subsection{Pulse profiles}

\begin{figure}
\centering 
\includegraphics[width=8.4cm]{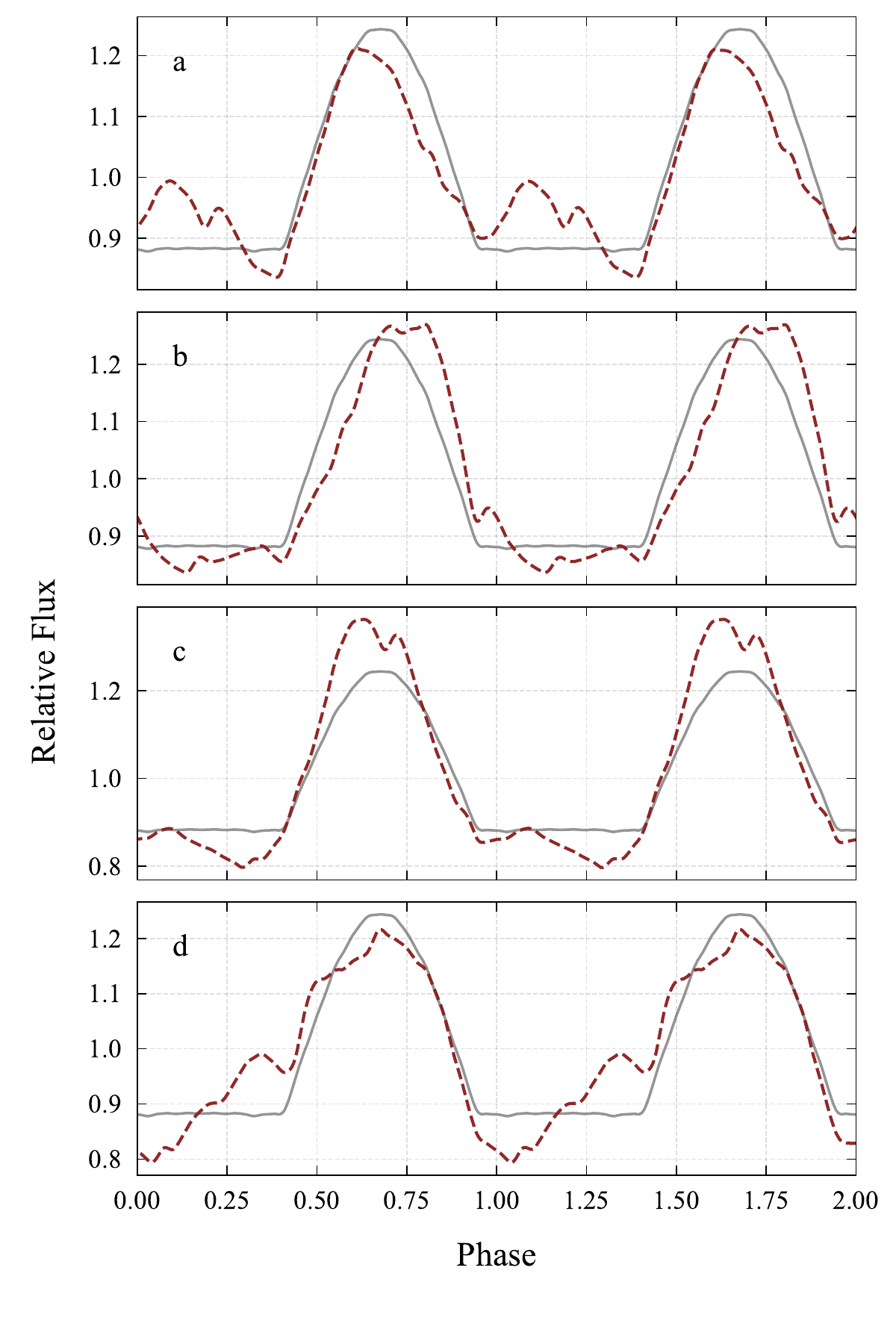}
\caption{
Pulse profiles calculated for a NS with fixed rotational geometry. 
The grey solid curve corresponds to the case of a constant mass accretion rate and is symmetric with respect to time reversal. 
Dashed red curves show pulse profiles obtained for a time-dependent accretion rate onto NS surface.
The temporal modulation of the accretion rate leads to a change in the pulse shape and the appearance of a pronounced asymmetry of the profile.
Parameters: 
$M=1.4\,M_\odot$,
$R=12\,{\rm km}$,
$P = 5\,{\rm s}$,
$R_{\rm m}=2\times 10^8\,{\rm cm}$,
$\alpha=0.6\,{\rm rad}$,
$\beta=0.05\,{\rm rad}$, 
$i = \pi/4$.
Photons are emitted by compact hot spots located at NS magnetic poles. The flux emitted by the spot is described by $F(\theta)\propto \cos\theta$.
}
\label{pic:sc_PP_varMdot}
\end{figure}

Fig.~\ref{pic:sc_PP_varMdot} illustrates the impact of a time-dependent mass accretion rate on the shape of pulse profiles. 
The accretion rate at the NS surface is modulated periodically due to the geometrical interaction between the rotating magnetosphere and the accretion flow according to red curve in Fig.~\ref{pic:flow_0}. 
For a given set of NS parameters, we compute the instantaneous emission from the surface and construct the corresponding pulse profile accounting for gravitational light bending (see Appendix~\ref{sec:LightBending}).

The solid black curve in Fig.~\ref{pic:sc_PP_varMdot} shows the pulse profile obtained under the assumption of a constant accretion rate to the poles of a NS. 
In this case, the profile is symmetric with respect to time reversal, as expected for a stationary beam pattern and purely geometrical modulation of the viewing angle.
The red curves correspond to the case of a variable mass accretion rate. 
Although the rotational parameters (inclination $i$ and magnetic obliquity $\alpha$) are kept fixed, the temporal modulation of the accretion rate introduces an additional degree of freedom in the problem. 
Specifically, the observed flux at a given rotational phase depends not only on the instantaneous viewing geometry but also on the phase of the accretion-rate modulation.
As a result, the observed pulse profile depends not only on $i$ and $\alpha$, but also on an additional azimuthal phase specifying the orientation of the observer relative to the pattern of accretion rate variability (see dashed lines in different panels of Fig.~\ref{pic:sc_PP_varMdot} that correspond to different values of the azimuthal angle).

Two main effects are clearly seen. 
First, the pulse profile is modified in shape compared to the stationary case. 
Secondly, the profile loses its symmetry with respect to time reversal. 
This loss of symmetry is a direct consequence of the time dependence of the emission pattern.
The flux at a given rotational phase is no longer uniquely determined by geometry, but also depends on the temporal evolution of the accretion rate.

Thus, even purely geometrical modulation of the accretion flow can lead to significant distortions of pulse profiles and naturally produce asymmetric light curves without invoking non-antipodal emission regions or complex magnetic-field configurations.
A quantitative comparison of the time-reversal asymmetry of the model profiles with the observed pulse profile of Cen~X-3 is presented in Appendix~\ref{App:assymetry}.

\section{Discussion}

\subsection{Physical motivation for spin--disc misalignment}

The physical motivation for considering a non-zero spin--disc misalignment deserves a brief comment.
For steady disc accretion, the NS spin axis is expected to align with the angular-momentum axis of the accretion flow on the spin-up time-scale \citep{2021MNRAS.505.1775B}. 
In this sense, a finite misalignment is unlikely to be a long-lived stationary configuration in persistently accreting systems. 
Be/XRPs, however, are often transient sources in which most of the angular momentum is transferred during relatively short outbursts. 
As a result, the effective alignment time-scale can be much longer than the instantaneous spin-up time inferred for continuous accretion. 
Moreover, the direction of the angular momentum supplied to the NS may vary between different accretion episodes. 
Numerical studies of Be/X-ray binaries indicate that the Be decretion disc can be substantially misaligned with the orbital plane and that this misalignment can strongly affect the geometry and efficiency of mass transfer during outbursts \citep{2013PASJ...65...41O}. Indeed, precession of the Be decretion disc has been observed in several transient XRPs \citep[see, e.g.,][]{2017MNRAS.471.1553K, 2025ApJ...984...66H}.
Under such conditions, a small residual angle between the stellar spin axis and the normal to a temporary accretion disc appears physically plausible, even if long-term evolution tends to drive the system towards alignment.
Recent X-ray and optical polarimetric observations provide direct observational support for such configurations. 
In particular, in the first Galactic ULX pulsar Swift J0243.6+6124, polarimetric analysis indicates that the NS spin axis is likely misaligned with the binary orbital axis by an angle of order several tens of degrees \citep{2024A&A...691A.123P}, demonstrating that appreciable spin–disc/orbit misalignment may occur in XRPs.
A related indication has recently been reported for the ULX pulsar M82~X-2 from XRISM  \citep{2025PASJ...77S...1T} phase-resolved spectroscopy of the Fe~K-shell band. 
\citet{2026arXiv260517786K} found pulse-phase variability of the Fe~K$\alpha$ emission and discussed a geometry in which the line-emitting accretion flow is not simply aligned with the binary/orbital configuration. 
Although the interpretation is not unique, this result provides additional motivation for considering spin–disc misalignment in ULX pulsars.


An additional motivation is provided by the observed superorbital variability in a number of XRPs, including Her~X-1, SMC~X-1 and LMC~X-4, which is commonly interpreted in terms of a warped and precessing accretion disc \citep{1976ApJ...209..562G,1976ApJ...210L.133D,1998ApJ...502..253W,2001MNRAS.320..485O,2003MNRAS.339..447C,2008NewAR..51..768C}. 
In some cases, free precession of the NS itself has also been discussed as a possible origin of the superorbital variability \citep{2009A&A...494.1025S,2013MNRAS.435.1147P,2024NatAs...8.1047H}.
In both scenarios, the mutual orientation of the disc and the stellar rotation axis changes with time, implying the presence of a non-zero spin--disc misalignment at least during part of the cycle. 


\subsection{Relation to global 3D MHD simulations}

The geometrical origin of the effect considered here is related to the behaviour seen in the global 3D MHD simulations of \citet{2021MNRAS.506..372R}. 
In their tilted-rotator models, the angle between the magnetic moment and the disc normal changed during the stellar rotation cycle, and episodes of larger magnetosphere--disc inclination were associated with more efficient funnel accretion. 
This provides an important physical precedent for the mechanism considered in the present work.

There is, however, an essential difference. 
In \citet{2021MNRAS.506..372R} the accretion rate variability was produced in a fully dynamical disc--magnetosphere system. 
The inner disc was warped, tilted, thickened and precessing, the magnetic field was distorted by the flow, and the accretion rate evolved together with the global disc structure. 
As a result, the signal was quasi-periodic and its amplitude changed with time. Such variability would not necessarily map coherently onto the X-ray pulse profile.

In the present work we consider the complementary limit. 
The disc is treated as a steady external mass reservoir and the magnetosphere is represented by a prescribed rotating dipole. 
This simplified setup removes the long-term MHD evolution of the disc and allows us to isolate the geometrical phase dependence of mass loading. 
We find that, in the large-magnetosphere regime, this geometrical effect produces a coherent modulation locked to the stellar spin period. 
This is the key new ingredient for pulse-profile modelling: the luminosity of the magnetic poles becomes phase-dependent in a reproducible way, rather than merely fluctuating quasi-periodically.

\subsection{Implications for pulse-profile modelling}

Our results suggest that part of the observed diversity and asymmetry of XRP pulse profiles may be explained without invoking complex magnetic field configurations or strongly non-antipodal emission regions \citep{1996ApJ...467..794K,2010A&A...517A...8S}. 
Instead, intrinsically time-dependent accretion driven by geometrical effects in the disc--magnetosphere interaction can naturally produce asymmetric and phase-dependent pulse shapes.
In addition, variations of the mass accretion rate are expected to affect not only the overall luminosity but also the structure of the emission region and the corresponding beam pattern of the X-ray radiation, that are expected to be mass accretion rate dependent \citep{1976MNRAS.175..395B}.
As a result, the temporal modulation of the accretion rate may introduce additional variability of the beam pattern itself, further modifying the pulse profiles.
This effect provides an additional source of complexity in pulse profile modelling, as the observed flux depends on both geometrical factors and the instantaneous accretion rate.

{
An interesting observational consequence of this picture may be relevant to transient Be/XRPs.
In these systems, the orientation of the temporary accretion disc may vary from one outburst to another because the Be decretion disc can be warped, tilted, or precessing \citep{2011MNRAS.416.2827M,2024MNRAS.528L..59M}.
The spin--disc misalignment angle may therefore also differ between accretion episodes.
Our model predicts that X-ray pulse profiles can change even at comparable luminosities if the geometry of the disc--magnetosphere interaction is different.
An observational example showing that comparable luminosities do not necessarily correspond to identical pulse profiles is provided by A~0535+26.
During its 2005 outburst, the energy-dependent pulse profiles observed during a short pre-outburst flare differed from those measured close to the maximum of the main outburst, although the two states reached comparable X-ray luminosities \citep{2008A&A...480L..17C}.
The same observations also revealed differences in the cyclotron-line energy.
These changes were interpreted in terms of a magnetospheric instability and a corresponding change in the field lines guiding the accretion flow.
Therefore, this particular case cannot be uniquely attributed to a change in the orientation of the accretion disc.
Nevertheless, it illustrates the more general point that the X-ray luminosity alone does not uniquely determine the accretion geometry or the pulse-profile morphology.
In transient Be/X-ray pulsars, variations in the orientation and structure of the temporary accretion disc may provide an additional source of such pulse-profile diversity.
}

\subsection{ULXs and pulsations in the presence of strong geometric beaming}

Strong geometric beaming is often invoked in the context of super-critical accretion onto magnetized NSs, where radiation escapes through a collimated funnel \citep{2009MNRAS.393L..41K,2017MNRAS.468L..59K}. 
In this regime, it has been argued that geometric beaming tends to suppress or strongly reduce pulsations (see, e.g., \citealt{2021MNRAS.501.2424M,2023MNRAS.518.5457M}).

This conclusion, however, relies on the implicit assumption that the mass accretion rate onto the NS surface is stationary and that pulsations arise solely due to the lighthouse effect associated with NS rotation. 
In this case, strong beaming can indeed smear out pulsations when emission from different rotational phases is redistributed over a narrow range of observer directions.

The situation is qualitatively different if the mass accretion rate itself is periodically modulated, as demonstrated in this paper. 
In this case, the total luminosity integrated over solid angle becomes an explicitly time-dependent quantity, varying on the spin period. 
As a result, pulsations are imprinted directly in the energy release, rather than being solely a geometric effect of changing viewing angle.

Under these conditions, geometric beaming does not suppress pulsations in a general sense. 
Instead, the beaming pattern redistributes a time-dependent luminosity over the sky, preserving the variability unless strong temporal smearing occurs within the emitting structure. 
Such smearing can become important only if the characteristic photon escape time from the beaming region is comparable to or exceeds the NS spin period (see Section~3.3 in \citealt{2021MNRAS.501.2424M}).

Therefore, in systems where the accretion rate onto the NS surface is intrinsically modulated, strong geometric beaming is compatible with pronounced pulsations.

Observationally, pulse profiles of ULX pulsars are typically broad and nearly sinusoidal, corresponding to a single-peaked modulation over the spin cycle (e.g., \citealt{2016ApJ...831L..14F,2017MNRAS.466L..48I,2017ARA&A..55..303K,2019MNRAS.488L..35S}). 
Such profiles are naturally expected if the luminosity itself is modulated, while the beaming pattern redistributes the emission over a finite angular range and introduces additional temporal smearing due to photon propagation within the funnel.

\subsection{Spectral signatures of accretion-rate modulation}

{Variability of the mass accretion rate can affect a broad range of spectral components in accreting XRPs. 
In particular, the energy and shape of cyclotron resonant scattering features are known to depend on pulse phase, which is commonly interpreted in terms of changes in the viewing geometry and beam pattern over the spin cycle \citep{2009A&A...498..825F,2015MNRAS.448.2175L}. 
In this context, the mechanism considered here may provide an additional contribution, because a periodic modulation of the accretion rate is expected to modify not only the luminosity, but also the structure of the emitting region and the beam pattern. 
As a result, the cyclotron line parameters may acquire an additional variability component linked to the intrinsic luminosity dependence.}

A similar effect can be expected for the continuum spectrum. 
The pulse phase dependence of the continuum shape (e.g. changes in the photon index and cutoff energy) is often interpreted purely as a consequence of viewing the emission region under different angles and, correspondingly, sampling regions of different optical depth and temperature. 
However, the situation may be more complex: if the mass-accretion rate itself varies with rotational phase, the physical conditions in the emitting region (density, temperature, optical depth, even geometrical structure) can change systematically over the spin cycle. 
This would naturally lead to continuum-shape variations that are not solely geometric in origin, but also directly driven by intrinsic accretion-rate modulation.

An analogous consequence arises for fluorescent iron K{\ensuremath{\alpha}} emission produced by reprocessing of the primary radiation in relatively cold material near the NS or the accretion disc. In this case, a periodically modulated accretion rate would lead to a corresponding modulation of the illuminating flux and may therefore produce pulse-phase variability of the iron line. Such behaviour has been reported in several XRPs \citep{1994ApJ...437..449C,2017AstL...43..175S,2018MNRAS.480.4746L}, and in this context the mechanism considered here provides a natural way to link the variability of the reflected emission to the intrinsic modulation of the accretion flow.
A recent XRISM \citep{2025PASJ...77S...1T} study of the ULX pulsar M82~X-2 reported pulse-phase variability of the Fe~K$\alpha$ emission together with evidence that the broadened component of the line is associated with the pulsating source itself \citep{2026arXiv260517786K}. 
While the detailed origin of the line remains uncertain, these observations are qualitatively consistent with a periodically varying illumination pattern and phase-dependent reprocessing in the inner accretion flow expected in the presence of spin–disc misalignment.

\section{Summary}

We have investigated the effect of a geometrical misalignment between the NS spin axis and the normal to the accretion disc on the mass accretion rate onto the stellar surface and the resulting X-ray pulse profiles. Our main results can be summarised as follows:

\begin{itemize}

\item 
We show that even for a stationary mass supply through the accretion disc, a non-zero spin--disc misalignment can convert steady disc accretion into a coherent spin-periodic modulation of the mass accretion rate onto the NS surface. 
This effect is purely geometrical and arises from the rotation of the misaligned magnetosphere relative to the disc. It isolates, in the large-magnetosphere regime relevant to classical XRPs, the geometrical tendency seen in the global 3D MHD simulations of \citet{2021MNRAS.506..372R}, where related accretion-rate variations appeared in a more complex, quasi-periodic disc--magnetosphere evolution.

\item The amplitude and shape of the accretion-rate variability depend on the system geometry, in particular on the angle between the spin axis and the disc normal, and on the relation between the stellar spin period and the characteristic propagation time of matter through the magnetosphere.

\item We find that a time-dependent accretion rate leads to noticeable modifications of X-ray pulse profiles even if all global parameters of the NS (mass, radius, rotational geometry) are fixed.

\item In particular, variability of the accretion rate naturally produces asymmetric pulse profiles by breaking the time-reversal symmetry expected for stationary emission patterns.

\item The resulting pulse profiles depend not only on the geometrical parameters of the system, but also on an additional azimuthal phase associated with the accretion-rate modulation, introducing an extra degree of freedom in modelling XRP light curves.

\end{itemize}

These results show that geometrical modulation of the accretion flow provides a natural mechanism contributing to the diversity and asymmetry of observed XRP pulse profiles, and must be taken into account when interpreting various phase-resolved properties of these sources (e.g. phase-resolved spectroscopy). The effect is expected to be most directly relevant when the modulation remains phase-locked to the stellar rotation rather than being dominated by stochastic or quasi-periodic variability of the disc--magnetosphere system.



\section*{Acknowledgements}

This research was supported by the International Space Science Institute (ISSI) in Bern, through the International Team project 25-657 `Polarimetric Insights into Extreme Magnetism' and the  Research Council of Finland Centre of Excellence in Neutron-Star Physics (grant 374064).
The authors thank Prof. Marina Romanova for valuable comments that helped to improve the manuscript.

\section*{Data availability}

The calculations presented in this paper were performed using a private code developed and owned by the corresponding author, please contact him for any requests/questions about it. All the data appearing in the figures are available upon request.


\appendix

\section{Numerical model}
\label{app:NumMod}

Following \citealt{2024MNRAS.530..730M,2026arXiv260710342M}, we simulate the plasma dynamics by following quasi-particles that represent equal-mass elements of the accretion stream. 
The code follows only the magnetically channelled component of the flow, which is assumed to move strictly along prescribed dipolar magnetic field lines defined by equation~(\ref{eq:dip_fl}). 
The disc--magnetosphere coupling itself is not solved self-consistently.
Since the numerical scheme itself has been described in detail in \citealt{2024MNRAS.530..730M}, here we summarize only those aspects that are important for the present problem.

\begin{figure}
\centering 
\includegraphics[width=8.4cm]{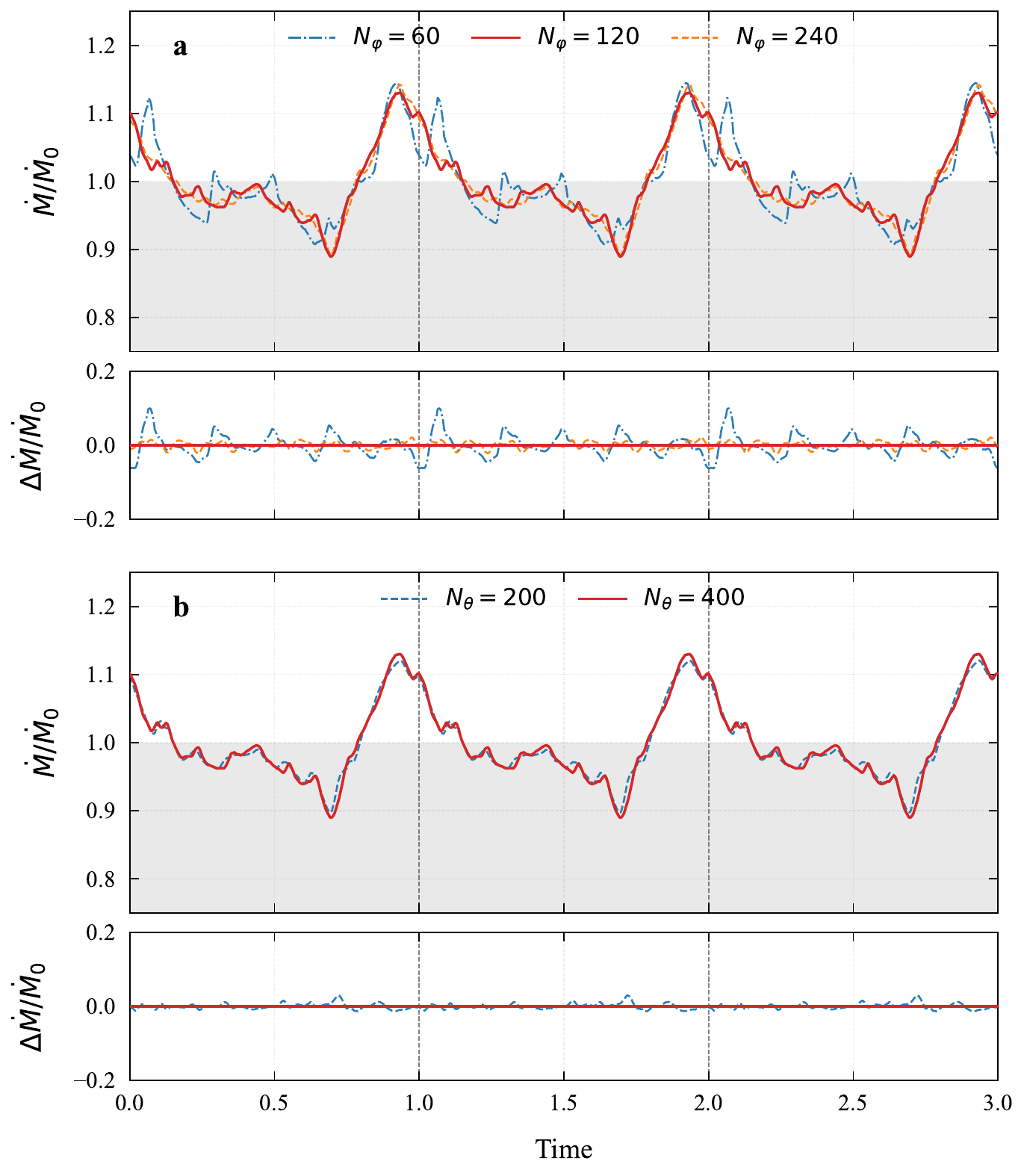}
\caption{
{
Numerical convergence tests for the mass accretion rate at the neutron star surface. 
The upper panels show the normalized mass accretion rate as a function of rotational phase, while the lower panels show deviations from the reference model. 
Panel (a) compares calculations performed with different azimuthal resolutions, $N_{\varphi}=60$, 120, and 240, for a fixed polar resolution $N_{\theta}=400$. 
Panel (b) compares calculations with different polar resolutions, $N_{\theta}=200$ and 400, for a fixed azimuthal resolution $N_{\varphi}=120$. 
In both cases, the reference model ($N_{\theta}=400$, $N_{\varphi}=120$) is shown by the red curve. The calculations were performed for the fiducial model with $\alpha=0.6$ rad, $\beta=0.05$ rad, $P=5$ s, $R_{\rm m}=2\times10^{8}\,$cm, and $v_{\rm i}=10^{7}\,{\rm cm\,s^{-1}}$. 
The lower panels demonstrate that the numerical differences remain small over the entire rotational cycle, confirming that the adopted spatial resolution is sufficient to reproduce the large-scale modulation of the mass accretion rate.}
}
\label{pic:sc_flow_comp}
\end{figure}

The geometry of the model is specified by two angles. The angle between the stellar spin axis and the magnetic dipole axis is denoted by $\alpha$, while the angle between the spin axis and the disc normal is denoted by $\beta$. The case $\beta=0$ corresponds to the standard configuration in which the spin axis is perpendicular to the disc plane. In the present work, the main interest is in the case $\beta\neq 0$, for which the orientation of the dipole with respect to the disc changes with rotational phase.

For each azimuthal sector, the code determines the instantaneous latitude $\lambda_{\rm d}(\varphi,t)$ at which the rotating dipolar field intersects the disc plane. 
Matter is injected in a narrow band around this latitude. Its angular half-width is estimated as
\beq
\delta\lambda \sim \frac{H_{\rm d}}{R_{\rm m}},
\eeq
where $H_{\rm d}$ is the geometrical half-thickness of the disc at the magnetospheric radius. For the gas-pressure-dominated thin discs considered here, $H_{\rm d}/R_{\rm m}\sim10^{-2}\ll1$, so the injection region occupies only a narrow range of magnetic latitude. In this way, even for a constant mass supply through the disc, the mass loading of individual field lines becomes explicitly phase-dependent when $\beta\neq0$.

The interval between the disc-interaction region and the stellar surface is divided into $N_\lambda$ segments in latitude and $N_\varphi$ zones in azimuth. 
In addition, the flow may in general be split into a finite number of layers across its thickness. 
In the calculations presented here, however, we use a single layer. 
This approximation is appropriate for the relatively low accretion rates ($\dot{M}\lesssim 10^{18}\,{\rm g\,s^{-1}}$) considered in this work, for which the flow is expected to remain optically thin across the magnetic field and significant transverse velocity gradients are not expected.
Thus, the magnetosphere is represented by a three-dimensional grid in $(\lambda,\varphi,{\rm layer})$.
For the calculations presented here, we adopt $N_\lambda=400$ and $N_\varphi=120$.
Each run follows the trajectories of injected particles over a physical time of several spin periods, using adaptive time steps to resolve the fast motion near the stellar surface.

We tested the sensitivity of the results to the spatial resolution by varying independently the numbers of grid cells along the magnetic field lines and in azimuth. 
The results of these tests are shown in Fig.~\ref{pic:sc_flow_comp}. 
Increasing the azimuthal resolution from $N_{\varphi}=120$ to $N_{\varphi}=240$ produces only minor changes in the large-scale modulation of the mass accretion rate, while the calculation with $N_{\varphi}=60$ reproduces the same overall profile with somewhat larger local deviations. 
Similarly, reducing the resolution along the field lines from $N_{\lambda}=400$ to $N_{\lambda}=200$ does not significantly affect the shape or amplitude of the modulation. We therefore conclude that the adopted resolution, $N_{\lambda}=400$ and $N_{\varphi}=120$, is sufficient for the results discussed in this work.
The remaining small-scale step-like features are associated primarily with the finite azimuthal discretisation and do not affect the large-scale
spin-periodic variability.

At every time step, the following procedure is applied:
\begin{enumerate}[leftmargin=*]
\item 
The mass supply through the disc, $\dot M_{\rm in}$, is specified at the inner boundary. 
In the baseline calculations discussed in this paper, $\dot M_{\rm in}$ is constant in time. 
The stochastic input fluctuations considered in \citet{2024MNRAS.530..730M} can be included if needed, but they are not required for the effect studied here.
\item 
A set of $N_{\rm p}$ particles is added to each azimuthal cell at the disc boundary.  
The mass assigned to each particle at time $t_i$ is
\beq
m_{{\rm p},j}=\frac{\dot{M}_{\rm in}\,\Delta t_i}{N_{\rm p} N_\varphi},
\eeq
or its obvious generalization when several flow layers are used.
The injection positions are randomized within the local disc-interaction region:
\beq
\lambda_j^{\rm (ini)} = \lambda_{\rm d}(\varphi_j,t_i) \pm \delta\lambda + X_j\Delta\lambda, 
\eeq
where $X_j\in (0,1)$ is a random deviate and the sign determines the initial direction of motion towards one of the magnetic hemispheres.  
The initial velocity along the field line is treated as a free parameter, motivated by the thermal speed of protons near $R_{\rm m}$:
\beq
v_{\rm ini}\sim v_{\rm p}\approx 
3\times 10^7\,T_{\rm keV}^{1/2}\,{\rm cm\,s^{-1}},
\eeq
and in practice we use $v_{\rm ini}\sim 10^7\,{\rm cm\,s^{-1}}$.
\item 
Knowing the positions and masses of particles, we evaluate the local mass and momentum content of the flow in each grid cell and reconstruct the corresponding magnetospheric maps.
\item 
The time step $\Delta t_i$ is adjusted dynamically to ensure that particles do not cross more than one grid cell per update:
\beq\label{eq:Delta_t_new}
\Delta t_i=\min\left\{
\min\limits_{j}\left[\frac{1}{C}\frac{(\d x/\d\lambda)\,\Delta\lambda_j}{|v_j(t_i)|}\right],
\Delta t_{\rm max}
\right\},
\eeq
where $C$ is a numerical safety factor of order unity.
\item 
The acceleration of each particle along the field line is computed from the projection of gravity and centrifugal force on to the magnetic-field direction. The radiative force can be included in the general version of the code, but it is neglected in the calculations presented in this paper. The Coriolis force does not contribute to the strictly field-aligned one-dimensional motion and is therefore omitted.
\item 
Particle velocities and positions are updated according to
\beq
v_{j}(t_{i+1})=v_{j}(t_{i})+ a_j(t_i)\Delta t_i,
\eeq
\beq
\lambda_{j}(t_{i+1}) = \lambda_j(t_i) + \Delta\lambda_j(t_i),
\eeq
with $\Delta\lambda_j$ obtained from the field-line geometry.
\item 
Particles reaching the stellar surface are removed from the list and contribute to the instantaneous accretion rate at one of the magnetic poles:
\beq\label{eq:dot_m_new}  
\dot{M}_k (t_i)= \frac{\sum_{j} m^{(k)}_{{\rm p},j}}{\Delta t_i}, 
\eeq
where $k=1,2$ indexes the poles.  
Particles that return to the disc-interaction region are not followed inside the disc. Instead, they are redistributed over magnetic meridians intersecting the disc and re-injected into the magnetosphere with the prescribed initial velocity. This prescription is intended to mimic matter redistribution near the disc--magnetosphere boundary while keeping the model computationally tractable.
\end{enumerate}

In this way, the evolving mass distribution over the magnetosphere is updated self-consistently together with the corresponding force balance. The key point for the present work is that the outer mass supply through the disc can remain strictly stationary, whereas the accretion rate on to the NS surface may nevertheless become periodic because the rotating misaligned dipole loads different field lines differently over the spin cycle.

\section{Modelling of stochastic mass accretion rate and power spectra}
\label{sec:AppPDS}

In order to model the variability of the mass accretion rate supplied by the disc, we follow the numerical approach described in our previous work \citep{2024MNRAS.530..730M}. 
Here we briefly summarise the key ingredients relevant for the present study.

We assume that the mass accretion rate at the inner disc radius exhibits stochastic variability characterised by a prescribed power density spectrum (PDS). 
The time series of the accretion rate is generated using the algorithm of \citet{1995A&A...300..707T}, adopting a power-law PDS of the form $P(f) \propto f^{q}$ within a finite frequency range. 
In the simulations presented here, we take $q \simeq -2$.

The generated time series $\dot{M}_{\rm d}(t)$ is then used as an outer boundary condition for the magnetospheric flow. 
The propagation through the magnetosphere acts as a non-linear filter of the input variability. 
To quantify the resulting variability at the NS surface, we construct the time series of the mass accretion rate onto the magnetic poles, $\dot{M}_{\ast}(t)$, directly from the numerical simulations.

The power density spectrum of the accretion rate is calculated via the discrete Fourier transform,
\beq
\dot{M}(f) = \Delta t \sum_{j=1}^{N_t} \dot{M}(t_j)\, e^{2\pi i f t_j},
\eeq
with the normalisation
\beq
{\rm PDS}(f) = \frac{2}{M_{\rm acc}} \left| \dot{M}(f) \right|^2,
\eeq
where $M_{\rm acc}$ is the total accreted mass over the considered time interval.

This procedure allows us to directly compare the variability properties of the accretion flow at the inner disc radius and at the NS surface, and to investigate how the magnetosphere modifies the broadband noise and introduces additional variability components.

\section{Light bending in Schwarzschild metric}
\label{sec:LightBending}

Photon propagation from the NS surface to a distant observer is treated in the Schwarzschild metric. 
This approximation is justified for XRPs with relatively long spin periods, where rotational corrections to the metric can be neglected.

The trajectory of a photon emitted at radius $r$ is determined by the geodesic equation, which can be written in terms of the impact parameter $b$ as
\beq
\left(\frac{1}{r^{2}}\frac{{\rm d}r}{{\rm d}\phi}\right)^{2}
= \frac{1}{b^{2}} - \frac{1}{r^{2}}\left(1 - \frac{2GM}{rc^{2}}\right).
\eeq
For practical calculations, it is convenient to relate the emission angle $\alpha$ (measured with respect to the local normal at the stellar surface) to the angle $\psi$ between the photon direction at infinity and the radial direction. 
This mapping fully determines the observed flux for a given surface emission pattern.

The observed specific flux is obtained by integrating contributions from visible surface elements, taking into account gravitational redshift and solid-angle transformation. In particular, photon energies are reduced by a factor
\beq
(1 - R_{\rm S}/R)^{1/2},
\eeq
where $R_{\rm S} = 2GM/c^{2}$ is the Schwarzschild radius.

In this work, photon trajectories are computed using a standard ray-tracing procedure in Schwarzschild geometry. 
The angular distribution of the observed flux is constructed by mapping emission directions at the stellar surface to directions at infinity. 
We follow the implementation described in \citealt{2024MNRAS.527.5374M} and apply it to the case of sub-critical XRPs, assuming emission from localized hotspots with a prescribed beam pattern.

\section{Quantifying the time-reversal asymmetry of pulse profiles}
\label{App:assymetry}

Visual inspection is often sufficient to determine whether a pulse profile is approximately symmetric with respect to time reversal. 
However, such comparisons are inherently subjective. 
To facilitate a quantitative comparison between different models and observations, we characterize each pulse profile by two simple dimensionless measures of time-reversal asymmetry.

Let $F_i$ be the pulse profile sampled at uniformly spaced rotational phases over one spin period. 
The corresponding time-reversed profile is obtained by reversing the phase order. 
Because the symmetry axis is not known a priori, the reversed profile is allowed to undergo an arbitrary cyclic phase shift before the comparison. 
The asymmetry measure is then defined as
\beq
A_{\rm direct}
=
\min_{\Delta}
\left[
\frac{
\sum_i
\left(
F_i-
F^{\rm rev}_{i,\Delta}
\right)^2
}{
\sum_i
\left(
F_i-\langle F\rangle
\right)^2
}
\right]^{1/2},
\eeq
where $\langle F\rangle$ is the mean flux over one spin period, $\langle F\rangle = N^{-1}\sum_i F_i$, and $\Delta$ denotes the
cyclic phase shift applied to the reversed profile.
The quantity $A_{\rm direct}$ is equal to zero for a perfectly time-reversal symmetric pulse profile and increases as the asymmetry becomes stronger. 
Since the comparison is performed point by point, it is sensitive to asymmetry on all spatial scales, including narrow peaks and other small-scale structures.

{
To quantify only the large-scale asymmetry of the pulse profile, we reconstruct the profile using only the first three Fourier harmonics,
\beq
F_{\rm low}(\phi) =
a_0 + \sum_{k=1}^{3}
\left[ a_k\cos(2\pi k\phi) + b_k\sin(2\pi k\phi) \right],
\eeq
and evaluate the same quantity,
\beq
A_{\rm low}
=
\min_{\Delta}
\left[
\frac{
\sum_i
\left(
F_{{\rm low},i}
-
F^{\rm rev}_{{\rm low},i,\Delta}
\right)^2
}{
\sum_i
\left(
F_{{\rm low},i}
-
\langle F_{\rm low}\rangle
\right)^2
}
\right]^{1/2},
\eeq
where $\langle F_{\rm low}\rangle$ is defined analogously as the phase-averaged value of the reconstructed profile.
Unlike $A_{\rm direct}$, the quantity $A_{\rm low}$ is largely insensitive to small-scale fluctuations and instead characterizes the asymmetry of the global pulse morphology.
}

\begin{table}
\centering
\caption{
Time-reversal asymmetry measures for the variable-accretion-rate pulse profiles shown by the red dashed curves in Fig.~\ref{pic:sc_PP_varMdot}. 
The four model profiles correspond to different azimuthal orientations of the observer.
The last row gives the values estimated from the observed 2--8~keV pulse profile of Cen~X-3 presented by \citet{2022ApJ...941L..14T}.
}
\label{tab:asymmetry}
\begin{tabular}{lcc}
\hline
Pulse profile &
$A_{\rm direct}$ &
$A_{\rm low}$ \\
\hline
Fig.~5(a) & 0.440 & 0.410 \\
Fig.~5(b) & 0.193 & 0.110 \\
Fig.~5(c) & 0.204 & 0.183 \\
Fig.~5(d) & 0.624 & 0.593 \\
Cen~X-3   & 0.630 & 0.614 \\
\hline
\end{tabular}
\end{table}

{
The asymmetry measures calculated for the variable-accretion-rate pulse profiles shown by the red dashed curves in Fig.~\ref{pic:sc_PP_varMdot} are summarized in Table~\ref{tab:asymmetry}. 
The four panels correspond to different azimuthal orientations of the observer relative to the phase-dependent accretion pattern. 
The resulting values span the ranges $A_{\rm direct}\simeq 0.19$--$0.62$ and $A_{\rm low}\simeq 0.11$--$0.59$. 
Thus, the apparent degree of time-reversal asymmetry depends strongly on the observer orientation, even though the rotational geometry and the accretion-rate modulation are the same in all four cases.
For comparison, we also evaluated the same quantities for the observed 2--8~keV pulse profile of Cen~X-3 shown in Fig.~9 of
\citet{2022ApJ...941L..14T}. 
The profile was digitized from the published figure and gives 
$A_{\rm direct}\simeq 0.63$ 
and
$A_{\rm low}\simeq 0.61$.
These values are close to those obtained for the model profile shown in panel (d) of Fig.~5, for which
$A_{\rm direct}=0.624$ 
and
$A_{\rm low}=0.593$.
The model profiles therefore cover degrees of time-reversal asymmetry comparable to that measured in Cen~X-3, although the asymmetry depends substantially on the orientation of the observer. 
This comparison shows that a phase-dependent accretion rate can produce not only a qualitative
distortion of the pulse profile, but also a level of large-scale asymmetry similar to that seen in an observed X-ray pulsar.
}

\bsp	
\label{lastpage}
\end{document}